\documentclass{aastex63}
\usepackage{bm}
\usepackage{ulem}
\usepackage{amsmath}
\usepackage{verbatim} 

\usepackage[utf8]{inputenc}
\usepackage[T1]{fontenc}

\usepackage[encapsulated]{CJK}

\shortauthors{Kino et al.}
\shorttitle{Magnetic field strengths in NGC~315}

\received{--}
\revised{--}
\accepted{--}

\begin{document}

\title{Mapping the distribution of the magnetic field strength along the NGC~315 jet}

\correspondingauthor{Motoki Kino}
\email{motoki.kino@gmail.com}

\author[0000-0002-2709-7338]{Motoki Kino}
\affil{Kogakuin University of Technology \& Engineering, Academic Support Center, 
 2665-1 Nakano-machi, Hachioji, Tokyo 192-0015, Japan}
\affil{National Astronomical Observatory of Japan, 2-21-1 Osawa, Mitaka, Tokyo 181-8588, Japan}

\author[0000-0002-7322-6436]{Hyunwook Ro} 
\affil{Korea Astronomy \& Space Science Institute, Daedeokdae-ro 776, Yuseong-gu, Daejeon 34055, Republic of Korea}

\author[0000-0001-9360-0846]{Masaaki Takahashi}
\affil{Department of Physics and Astronomy, Aichi University of Education, Kariya, Aichi 448-8542, Japan}

\author[0000-0001-8527-0496]{Tomohisa Kawashima}
\affil{Institute for Cosmic Ray Research, The University of Tokyo, 5-1-5 Kashiwanoha, Kashiwa, Chiba 277-8582, Japan}

\author[0000-0001-6558-9053]{Jongho Park}
\affil{School of Space Research, Kyung Hee University, 1732, Deogyeong-daero, Giheung-gu, Yongin-si, Gyeonggi-do 17104, Republic of Korea}

\affil{Institute of Astronomy and Astrophysics, Academia Sinica, P.O. Box 23-141, Taipei 10617, Taiwan}

\author[0000-0001-6906-772X]{Kazuhiro Hada}
\affil{Graduate School of Science, Nagoya City University, Yamanohata 1, Mizuho-cho, Mizuho-ku, Nagoya, 467-8501, Aichi, Japan}
\affil{Mizusawa VLBI Observatory, National Astronomical Observatory of Japan, 2-12 Hoshigaoka-cho, Mizusawa, Oshu, 023-0861, Iwate, Japan}
\affil{Astronomical Science Program, The Graduate University for Advanced Studies (SOKENDAI), 2-21 Osawa, Mitaka, 181-8588, Tokyo, Japan}

\author[0000-0001-6311-4345]{Yuzhu Cui}
\affil{Zhejiang Laboratory, Hangzhou, People’s Republic of China}

\begin{abstract}

We study magnetic field strengths along the jet in NGC~315.
First, we estimated the angular velocity of rotation in the jet magnetosphere by comparing the measured velocity profile of NGC~315 with the magneto-hydrodynamic jet model of proposed by Tomimatsu and Takahashi.
Similar to the case of M87, we find that the model can reproduce the logarithmic feature of the velocity profile and suggest a slowly rotating black hole magnetosphere for NGC~315. 
By substituting the estimated $\Omega_{F}$ into the jet power predicted by the Blandford-Znajek mechanism, we estimate the magnetic field strength near the event horizon of the central black hole
as $5\times 10^{3}~{\rm G}\lesssim B_{H}\lesssim 2\times 10^{4}~{\rm G}$.
We then estimate magnetic-field strengths along the jet by comparing the spectral index distribution obtained from VLBI observations with a synchrotron-emitting jet model. 
Then we constrain the magnetic field strength at a de-projected distance $z$ from the black hole to be in the range $0.06~{\rm G}\lesssim B(z)\lesssim 0.9~{\rm G}$ for $5.2 \times 10^{3}~r_{g}\lesssim z \lesssim 4.9 \times 10^{4}~r_{g}$, where $r_{g}$ represents the gravitational radius.
By combining the obtained field strengths at the event horizon and the downstream section of the jet, we find that the accretion flow at the jet base is consistent with a magnetically arrested disk (MAD). 
We discuss a comparison of the jet power and the magnetic flux anchored to the event horizon in NGC~315 and M87.

\end{abstract}

\section{Introduction}
\label{sec:intro}

Magnetic fields are believed to play an essential role in launching relativistic jets in active galactic nuclei (AGN) \citep[e.g.,][and references therein]{bbr84, blandford19}{}{}. The current leading scenario for jet launching mechanism is a magnetically driven jet ultimately powered by the rotational energy of the central black hole proposed by \citep[][hereafter BZ77]{bz77}.  
Therefore, testing BZ77 is crucial to understanding the mechanism of jet formation. One of the most important tasks for testing BZ77 is to reveal the strength of the magnetic field from the scale of the event horizon to the downstream of the jet and its radial distribution along the jet.
The nearby massive elliptical galaxy M87 provides an ideal laboratory for investigating properties of magnetic fields in great detail.
The first imaging  of M87's black hole shadow 
by Event Horizon Telescope (EHT) at 230~GHz has provided a giant leap in our understanding of black holes \citep[][]{EHT19_1,EHT19_2,EHT19_3,EHT19_4,EHT19_5,EHT19_6}.
The EHT further detected the linearly polarized emission with the spiraling structure, suggesting a signature of poloidal magnetic fields threading the accretion flow \citep[][]{EHTC7,EHTC8}.
The measurement of the jet velocity profile of the M87 jet
based on high-cadence  KaVA array observations at 22~GHz and 43~GHz 
were performed by \citet[][]{Park19}. 
The measured velocity field showed a clear  discrepancy with the velocity field predicted by general relativistic magnetohydrodynamics (GRMHD) simulations performed where the jet propagates long distances on the order of $10^{4-5}~r_{g}$ \citep[][]{McKinney06, chatterjee19} 
where $r_{g}\equiv GM_{\bullet}/c^{2}$ is the gravitational radius of the central black hole and 
$M_{\bullet}$ is the mass of the central supermassive black hole.

It posed the challenge of explaining the observed jet velocity profile with GRMHD simulations.
To mitigate the discrepancy between the observation and theoretical models in the jet velocity field, slowly rotating black hole magnetosphere (i.e., the small angular velocity of the magnetic field $\Omega_{F}$) has been proposed where the smaller $\Omega_{F}$ makes a light-cylinder radius larger, and it consequently pushes out a location of a starting point of the jet acceleration \citep[][]{kino22}.
\footnote{
Some may consider entrainment as a physical process to explain the slower velocity within $\sim 10^{3}~r_{g}$.
It could be possible to decelerate the flow with intense entrainment with surrounding ambient material after the jet's bulk acceleration reaches its maximum velocity $\Gamma_{max}$ at a distance of about $10^{2}~r_{g}$ from the central engine, as suggested in GRMHD simulations
\citep[e.g.,][Fig. 13]{chatterjee19}. The thermal energy generated during the interaction between the jet and surrounding medium may help re-accelerate the jet.
However, the numerical simulation by \citet{ricci23}, which investigated thermal acceleration, did not show sufficient deceleration, resulting in a significant overestimation the jet velocity within $\sim 10^{3}~r_{g}$. Moreover, such significant deceleration and re-acceleration do not produce a smooth velocity profile but rather a beat-like profile, which is not suggested in the velocity profile of NGC~315.
For these reasons, explaining the velocity profile through the entrainment phenomenon appears challenging.}
By using the estimated $\Omega_{F}$, one can 
make the order estimation of 
the magnetic field strength on the event horizon scale ($B_{H}$)
in M87 by assuming Blandford–Znajek process is in action.

However, our current understanding of the profile of the magnetic field strengths along jets in AGN is limited only for M87 
\citep[][and the references therein]{Ro23}, 
because of the limitations of the number of samples. For this, it generally requires both of
(1) largeness of angular size of a central black hole, and 
(2) measurement of a jet's velocity profile with enough range of distance in order to know jet's acceleration feature.
Jet acceleration features from non-relativistic to relativistic speed in  jet collimation regions have been
measured only in 
M87 \citep{kovalev07, ly07, acciari09, Asada14,Hada16, Hada17,Mertens16, walker18, Kim18, Park19},
Cygnus A \citep[][]{krichbaum98, bach03, boccardi16},
1H 0323+342 \citep[][]{hada18}, and 
NGC~315 \citep[][]{park21,ricci22}.
Unfortunately, low-luminosity AGNs having a large angular-size black hole at their center \citep[e.g.,][and references therein]{Ramakrishnan23} do not always have a clear jet feature 
and rather limited numbers of samples.

The nearby giant elliptical radio galaxy NGC~315
located at the redshift of 
$z_{\rm redshift} = 0.01648$ 
\citep{trager00}
is the suitable target for investigating
the jet acceleration mechanism,
because the proximity of distance 
and the large mass of the black hole 
enables us to resolve the innermost part of 
the jet base.
The mass of the central black hole in NGC~315 has been estimated
based on gas kinematics \citep[][]{noel-storr07}, 
the black-hole mass and stellar-velocity dispersion relation
($M_{\bullet}\--\sigma$ relation)
\citep[e.g.,][]{woo02, beifiori09, inayoshi20}, and 
CO observations with the Atacama Large Millimater/submillimeter Array (ALMA) \citep{boizelle21},
which lies in the range
$M_{\bullet}=(1.2-2.1)\times 10^{9}~M_{\odot}$.
Thanks to  its proximity and largeness of the central black hole,
one can explore the innermost part of the NGC 315 and
the measured velocity field of the NGC~315 jet 
clearly show the acceleration feature from non-relativistic speed to the relativistic one 
thanks to  its proximity and largeness of the 
central black hole \citep[][]{park21,ricci22}.
NGC~315 is also known one of the GeV $\gamma$-ray emitters.
Palomar spectroscopic survey of nearby galactic nuclei  make it suitable to the study of nearby AGN, especially low luminosity AGN
\citep{ho95,ho97,ho08}.
Of the 197 low luminosity AGN in the Palomar sample, only four sources were detected in $\gamma$-rays \citep{deMenezes20}.
These four sources correspond to misaligned Fanaroff Riley I 
(FR I) type radio galaxies, accompanying kpc-scale radio jets.
Two of them are NGC 315 and M87 and they have similar $M_{\bullet}$ and SED.
Towards understanding the jet launching mechanism, it is of great importance to investigate and compare the properties of magnetic fields in these jets.

In \S~2, we briefly overview the models to be utilized in the present work.
In \S~3, we apply these models to the NGC~315 jet and 
describe the obtained results.
In \S~4, we will compare the obtained physical properties of NGC~315 and M87.
In \S~5, we summarize our findings.
In this work, we assume the standard $\Lambda$-CDM cosmology with the  parameter values of $H_{0} = 67.4 \,{\rm km\,s^{-1}\,Mpc^{-1}}$, and 
$\Omega_{\rm m}=0.315$ \citep[see][]{Planck2020}.
Hereafter, we use the median value $M_{\bullet}=1.6\times 10^{9}~M_{\odot}$ for the black hole mass.
This corresponds to a linear 
scale of $0.33~{\rm pc/mas}$ 
or equivalently 
$1{\rm mas} = 4.57 \times 10^{3}~r_{g}$
at a distance of $D=72.1$~Mpc.
The Schwarzschild radius is given by $r_{s}=2~r_{g}$.

\section{Models}
\subsection{Dynamics: MHD jet model}
In this work, we utilize the MHD jet model proposed by \citet[][]{TT03} 
(hereafter TT03) in the framework of special relativity. 
In the black hole magnetosphere,
due to the balance between the gravitational force of the 
black hole and magneto-centrifugal force, a stagnation 
(a.k.a separation) surface is generated that separates the inflow and outflow regions
\citep[e.g.,][]{Takahashi90,McKinney06,Pu15,Pu20}.
Comparing semi-analytical approaches and GRMHD simulation approaches, the semi-analytical approaches have the following advantages.
The large spatial extent of the acceleration region
has posed a challenge for such calculations by GRMHD simulations
and they tend to be eventually
limited by computational costs and numerical dissipation 
\citep[e.g.,][for details]{McKinney06, Komissarov07}, 
while semi-analytic approaches are free from these concerns.
When discussing properties of axisymmetric and steady MHD flows in general, 
the magnetic field geometry should be consistent with Grad-Shafranov equation \citep[e.g.,][]{nitta91}, 
and the flow should be trans-fast-magnetosonic
\citep[][]{Takahashi90,takahashi02}.
However, it is technically difficult to obtain a solution satisfying both of these conditions 
\citep[e.g.,][for review]{Beskin10}.
TT03 model is the only semi-analytic solution to satisfy both of these conditions.
Given that a radial profile of the poloidal component of the jet's four-velocity ($u_{p}$) is
sensitive to the magnetic field geometry, 
we utilize TT03 model in this work.
TT03 model is applicable to the region outside the light cylinder ($r_{\rm lc}$). In general, the jet is accelerated outside the light cylinder.
\footnote{
Flow properties inside the light cylinder have been recently investigated 
by several authors
\citep[e.g.,][]{takahashi21, camilloni22, beskin23}. 
\citet{beskin23} suggests that
 a dense cylindrical core appears
 over long enough distances near the axis
 with its radius comparable to $r_{\rm lc}$.
\citet{camilloni22} showed that
in the force-free magnetosphere in Kerr geometry, there is a logarithmic term in the angular velocity.}

The outer boundary wall condition would be given by
a parabolic streamline along the poloidal magnetic field lines. 
Since multiple normalizations are conducted, it would be helpful to explicitly write down the boundary condition here.
The outer boundary wall shape is denoted as ($Z$, $R$)
in the cylindrical coordinate
and it satisfies the following relation:
\begin{align}\label{eq:boundary-wall}
\left(\frac{1}{{\cal E}}
\frac{Z}{r_{\rm lc}}\right) \theta_{0} = 
\left(\frac{1}{{\cal E}}\frac{R}{r_{\rm lc}}\right)^{q} 
(1\leq q \leq 2),
\end{align}
where the 
$\theta_{0}$ is the half-opening angle of
the jet at the inlet boundary 
and the magnetic flux function on the boundary wall 
satisfies 
$\Psi(z=Z,r=R) = \Psi_{0}$.
Note that
the case of $q=1$ corresponds to 
a conical boundary wall shape.
In this work, we will give the value of $q$ 
with reference to the overall results of the detailed 
VLBI observations in \S 4.

On the contrary to the case of varying ${\cal E}$,  
$\Omega_{F}$ does not 
alter the profile of $u_{p}$ itself.
As already shown, $\Omega_{F}$ is governed by
the light cylinder radius $r_{\rm lc}$
and it is given by
$\Omega_{F}= c/r_{\rm lc}$.
%
Slower rotation of $ \Omega_{F}^{-1}$ 
leads to more distant starting point of the jet acceleration from the central black hole.

In this work, a location of intersection between 
the boundary-wall ($\Psi=\Psi_{0}=1$ surface)
and the light-cylinder is important. 
Hereafter,
we denote the location as ($R_{0}$, $Z_{0}$).
By inserting $R_{0}=r_{\rm lc}$ at Eq.~(\ref{eq:boundary-wall}),
the location of $Z_{0}$, from which 
the jet acceleration starts \citep[see Figure 1 in][]{kino22}, 
is obtained as follows:
\begin{align}
Z_{0}=100~r_{g}
\left(\frac{\theta_{0}}{0.1}\right)^{-1}
\left(\frac{r_{\rm lc}}{10\,r_{g}}\right)
{\cal E}^{1-q}   .
\end{align}
We note that the geometrical factor $\theta_{0}$ also affects the location of $Z_{0}$.

\subsection{Nonthermal synchrotron emission model}
\label{sec:syn-model}

Following \citet{Ro23}, we introduce a spectral index model along the jet. The essence of the model can be explained as follows.
A change of a spectral index along the jet reflects a change 
in the nonthermal electrons' energy distribution function (eDF hereafter) along the jet. 
Spectral indices of nonthermal electrons ($p$) and nonthermal synchrotron emission ($\alpha$) are connected as $\alpha=(p+1)/2$
where the sign of the spectral index as $S_{\nu}\propto \nu^{+\alpha}$ is defined and $S_{\nu}$ presents synchrotron flux per unit frequency.
Time evolution of eDF is described by the continuity equation including nonthermal electron injection and energy losses.
(1) In the case of a constant and continuous injection of non-thermal electrons, the synchrotron spectrum has a break, and the spectral index steepens by $0.5$ before and after the break. 
(2) For instantaneous injection, the cutoff energy $\gamma_{\rm max}$ decreases with time. Above the cutoff energy, the synchrotron spectrum decreases exponentially.
 Neither (1) nor (2) alone can explain the properties of the observed spectral index of the M87 jet.
To overcome this problem, \citet{Ro23} proposed a model where the injection-rate function of nonthermal electrons has $z$-dependence written as
\begin{align}
    Q_{e,\rm inj}(\gamma_{e},z)= Q_{0} \gamma_{e}^{+p_{\rm inj}} 
    \left(\frac{z}{z_{0}}\right)^{-q_{\rm inj}}
\end{align}
where $Q_{0}$ and $p_{\rm inj}$ are a normalization factor
and a spectral index of injected nonthermal electrons, respectively. The index $q_{\rm inj}$  describes 
$z$-dependence of $Q_{e,\rm inj}$
\citep[see details][]{Ro23}.
By matching an observed synchrotron spectral index and a model-predicted one, one can constrain on strengths of magnetic fields along the jet ($B(z)$) and the nonthermal electron injection function.
The model generally describes that
the stronger magnetic field strength realizes
the closer distance for the steeping of the spectral index due to synchrotron cooling in the jet.
A smaller $q_{\rm inj}$ leads to a larger injection of the nonthermal electrons, which compensates for the reduction of the number of electrons due to the cooling at a certain  distance along the jet.
It realizes a suppression of the steepening of the spectral index (i.e., larger $\alpha$).

\section{Observational properties of NGC~315}

\subsection{The jet collimation profile}

The jet collimation profile of NGC~315 has been 
explored in previous works \citep[][]{park21,boccardi21}.
Transverse intensity profile along the jet axis
is obtained by the multi-frequency observations
and the jet width at each distance is obtained.
The obtained jet width profile
defined as $Z\propto R^{q}$ in the previous section
shows the parabollic shape characterized by
\begin{align}
\begin{cases}
q &= 1.72 ~ (z\lesssim 10^{5}~r_{g}) \\
q &= 0.86 ~ (z\gtrsim 10^{5}~r_{g})
\end{cases}
\end{align}
measured by \citep[][]{park21}.
In this work, we do not treat the rapidly expanding jet region (so-called geometrical flaring region; Canvin et al. 2005) at $>10^{5}~r_{g}$, because it is beyond the application of the TT03 model. 
Hereafter, we focus on the inner parabolic region of acceleration and collimation zone of the NGC~315 jet with the index 
$q=1.72$ during the application of
TT03 model to NGC~315.

\subsection{The jet base opening angle ($\theta_{0}$)}

Following the previous work of \citep[][in Figure 4]{ricci22} where the intrinsic half-opening angle has been measured, hereafter the full-opening angle is set as
\begin{align}
4^{\circ} \lesssim \theta_{0} \lesssim 6^{\circ} .
\end{align}
This value falls in the typical range of 
$\theta_{0}$ for various radio galaxies
\citep[e.g.,][]{pushkarev17}.
This $\theta_{0}$ value of NGC~315 satisfies
guarantees that $\theta_{0} \ll 1~{\rm radian}$.

\subsubsection{The poloidal velocity of the jet ($u_{p}$)}

The poloidal component of the jet velocity
$u_{p}=\Gamma v_{p}$ (four-velocity)
of NGC~315 was derived from the observed jet-to-counterjet intensity ratio ${\cal R}=I_{\rm jet}/I_{\rm cjet}=[(1+\beta\cos\theta_{\rm view})/(1+\beta\cos\theta_{\rm view})]^{2-\alpha}$
where 
$\beta=v_{p}/c$,
$\Gamma=1/\sqrt{1-\beta^{2}}$, and 
$\theta_{\rm view}$ are 
the jet velocity (three-velocity), 
the jet bulk Lorentz factor, and 
the angle between the jet axis and our line of sight,
respectively \citep{park21}.
Here it is assumed that 
the jet and the counterjet are intrinsically the same and the toroidal component of the jet velocity
is negligible.
Solving the degeneracy between $\beta$ and $\theta_{\rm view})$  was carefully done
by comparing the observed 
${\cal R}$ and $\alpha$ with the jet kinematics
obtained by the monitoring observations with KaVA array and the derived the angle
is $\theta_{\rm view}\approx 50^{\circ}$
\citep{park21}, which is consistent with
the independent estimation of the viewing angle 
on kilo-parsec scale jet \cite{LB14}.
Therefore, in the present work,
we use the data obtained in \citep{park21}.
Note that \citet{ricci22} also suggested a similar trend for
the jet velocity profile of NGC~315.
As was pointed out in \citet{park21},
the measured $u_{p}$ of NGC~315 shows
"the slow acceleration" as was firstly 
seen in the M87 jet \citep{Park19}.
\citet{park21} suggests that possible ingredients to make it happen may be a variety of magnetization degree at the jet base,
and/or interaction between the jet and ambient medium is suggested.
Recently \citet{kino22} proposed a new scenario where a slower angular velocity of the black hole magnetosphere $\Omega_{F}$ can mitigate the slow acceleration problem.
The maximum Lorentz factor of NGC~315 reaches about
\begin{align}
\Gamma_{\rm obs, max}\approx 3.7   ,
\end{align}
at $z \sim 7\times 10^{4}~r_{g}$.
The measured  $u_{p}(z)$ of NGC~315 shows the rapid deceleration 
at the outer region of $Z_{\rm brk}$ probably suggests
the significant interaction
between the jet and surrounding matter
that causes the observed deceleration
 FR I jets may decelerate 
 by entrainment of the surrounding matter 
 \citep[e.g.,][]{baan80,bicknell84,deyoung93,kawakatu09} 
 or by injection of mass lost by stars 
 within the jet volume 
 \citep[e.g.,][]{komissarov94,perucho14}.
Whichever case realizes, 
the region $z > Z_{\rm brk}$ is beyond the application of the TT03 model and thus we do not treat this decelerating region in this work.

\subsection{The jet power ($L_{j}$)}

It is generally difficult to constrain on 
the total power of AGN jets, 
 because of the existence of invisible components such as low-energy electrons/positrons and protons
 \citep[e.g.,][and references therein]{kino12,sikora20}.
Following the work of \citet{ricci22},
we set the allowed range of the total power of the jet ($L_{j}$) as follows:
\begin{align}
2 \times 10^{43}~{\rm erg~s^{-1}} \lesssim L_{j} \lesssim 1.4\times 10^{44}~{\rm erg~s^{-1}}.
\end{align}
The lower limit is estimated from the Bondi accretion rate \citep{nemmen15}, 
while the upper limit is derived from the empirical relationship between the radio core luminosity and $L_{j}$ using the empirical relationship between core luminosity and jet power \citep[e.g.,][]{Heinz07}.

\subsection{Mass accretion rate ($\dot{M}$)}\label{sec:mdot}

A mass accretion rate ($\dot{M}$) can be estimated 
in several ways.
Firstly, $\dot{M}$ can be estimated based on a bolometric luminosity ($L_{\rm bol}$)
by using the relation of $\dot{M} = L_{\rm bol}/\eta c^{2}$ where $\eta$ is the conversion efficiency of the mass accretion energy to $L_{\rm bol}$.
The bolometric luminosity in NGC~315 can be derived from the [O III] 5007 emission line together with the relation $L_{\rm bol} = 3500 L_{\rm [OIII]}$.  
The $L_{\rm bol}$ can be obtained also from a direct integration of the nucleus's SED. Both of them indicate the consistent value of  
$L_{\rm bol} = 2\times  10^{43}~{\rm erg s^{-1}}$ \citep{Gu07,ricci22}
which derives $\dot{M} \sim 10^{-4}~M_{\odot}~{\rm yr^{-1}}$
when assuming $\eta = 0.1$
\citep{ricci22}.
Secondly, $\dot{M}$ may be estimated 
when HI broad absorption line is detected
\citep{vanGorkom89}
that can be related to the fueling gas that is falling into the nucleus.
\citet{morganti09} estimated $\dot{M}$ in NGC~315
and it is suggested as
$10^{-4}~M_{\odot}~{\rm yr^{-1}}
 \lesssim \dot{M} \lesssim 
 10^{-3}~M_{\odot}~{\rm ~yr^{-1}}$
using the assumed infalling velocity of
$50\,{\rm km \,s^{-1}}$.
Other independent work of SED model fitting for the central region of 
NGC~315 also suggest a comparable value of 
$\dot{M} = 4 \times 10^{-4}~M_{\odot}~{\rm yr^{-1}}$
\citep{KT20}.
Therefore, the allowed range of $\dot{M}$ in this work is set as 
\begin{align}
10^{-4}~M_{\odot}~{\rm yr^{-1}}
 \lesssim \dot{M} \lesssim 
 10^{-3}~M_{\odot}~{\rm yr^{-1}},
\end{align}
in the following discussion.

In addition, \citet{morganti09} estimated $\dot{M}$ 
by using the empirical relationship between $L_{j}$ and Bondi accretion power \citep[e.g.,][]{Allen06, Barmaverde08}.
Although the estimated maximum mass accretion rate 
$\dot{M}_{\rm max}\sim 10^{-1}~M_{\odot}~
{\rm yr^{-1}}$
seems extremely large,
we will investigate this case as well as an additional one.

\section{Estimation of magnetic field strengths}

Here we show the estimated strengths of 
the magnetic fields along the NGC~315 jet 
by applying the above-explained models.

\subsection{Constraining $B_{H}$} 
\label{sec:BH}

\begin{figure} 
\includegraphics
[width=18cm]
{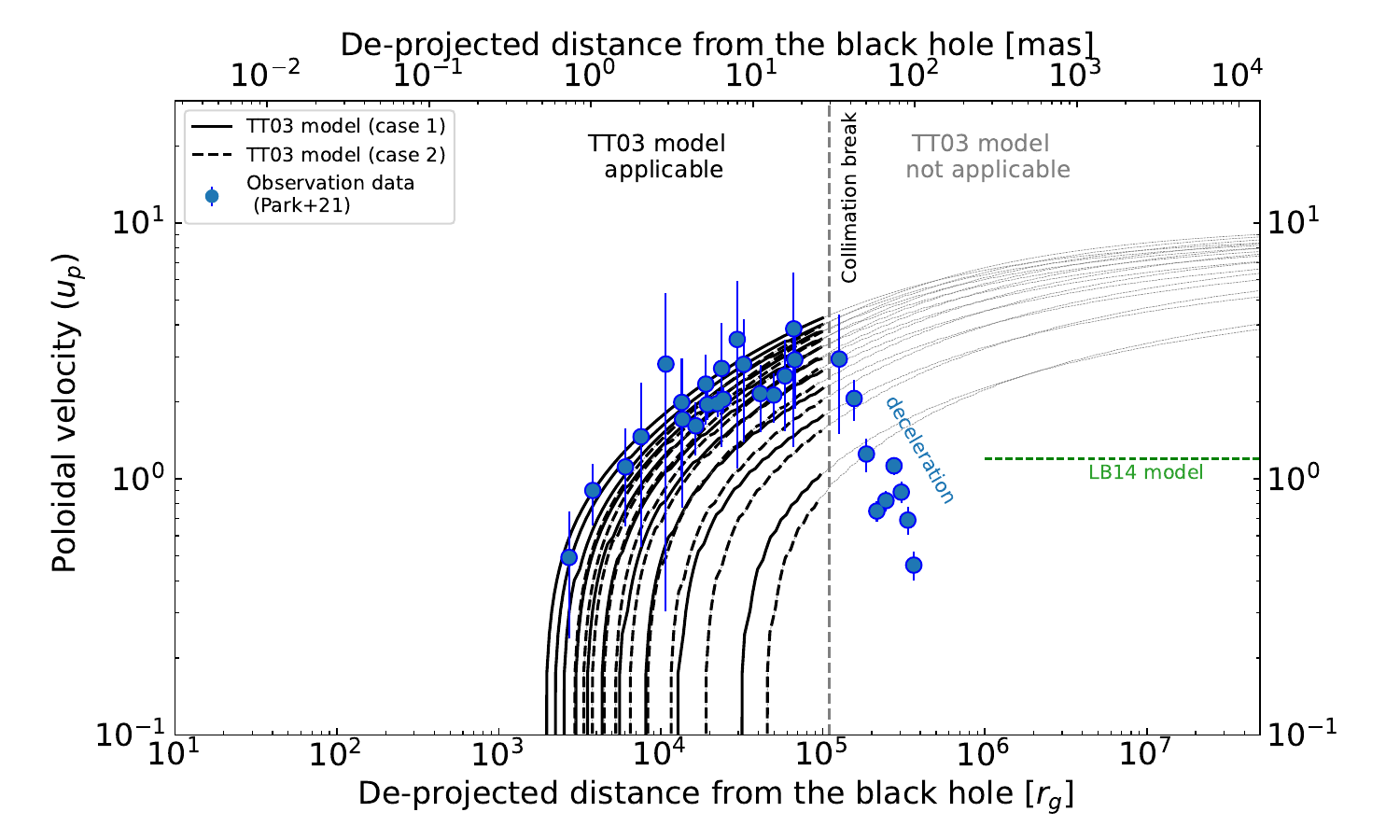}
\caption
{
The comparison of observed and model-predicted jet velocity profile.
The vertical axis presents the poloidal component of the jet's four-velocity denoted as $u_{p}=\Gamma v_{p}$.
The blue circles are the measured velocity profile of the NGC~315 jet 
\citep{park21}.
The solid curves show the velocity profile predicted by TT03 model.
The model parameters in case 1 are $\Omega_{F}=c/(500~r_{g})$ and ${\cal E}=8$,
while case 2 has $\Omega_{F}=c/(800~r_{g})$ and ${\cal E}=11$.
 The dashed vertical lines show the distance of the jet collimation break.
As a reference, the model-predicted
velocity fields on kilo-parsec scales by \citet{LB14} 
is shown (the green dashed line). }
\label{fig:NGC315up1}
\end{figure}

Figure \ref{fig:NGC315up1} presents the best-fit profile of the model-predicted $u_{p}$ overlaid to the VLBI-measured $u_{p}$.
As seen here, the jet flow is described as multiple streamlines of the plasma flow 
along the magnetic surfaces with their curvature
is calculated in TT03 model.
As emphasized in \citep[][]{kino22}, 
the key feature is the location of the starting point of the jet acceleration.
The allowed range of the angular velocity is 
\begin{align}
\frac{c}{800~r_{g}}\lesssim \Omega_{F}
\lesssim \frac{c}{500~r_{g}}
\end{align}
Errors bars for the innermost measurement of $u_{p}(z\sim  3 \times 10^{3}~r_{g})$, 
introduce uncertainties for the determination of 
the starting point of the jet acceleration
\footnote{
It is noteworthy that a recent work of \citet{ricci23} shows that if thermal energy in the jet is comparable to or exceeds magnetic energy, thermal acceleration becomes significant at parsec scales. 
If this is the case, then the discrepancy between the observed velocity field and the model-predicted one would get even larger.}.

In order to take into account such uncertainties, here we have conducted two cases of the fittings.
Case 1 fits the higher value of the data point 
that realizes a smaller width of $r_{\rm lc}$,
while Case 2 fits the lower value of the data point 
that realizes a larger width of $r_{\rm lc}$.
The resultant small difference in the starting point 
of the jet acceleration is seen in Figure~\ref{fig:NGC315up1}.
For both cases, we fit the observation data point with the fastest speed at $z\sim 7\times 10^{4}~r_{g}$.
Since the jet is only logarithmically accelerated,
Case 2 requires a larger ${\cal E}$ than Case 1.

The dashed vertical lines show the distance of the jet collimation break.
 The observation data show the  jet deceleration at a distance further away than
 the jet collimation break (i.e., $z>Z_{\rm brk}$).
 At the jet deceleration region, TT03 model is not applicable.
In addition,
it would be worth noting the Bondi radius 
$R_{\rm Bondi} = {2GM_{\bullet}/c_{s}^{2}}
=0.031~{\rm kpc}(kT_{e}/{\rm keV})^{-1}(M_{\bullet}/10^{9}M_{\odot})$ 
within which the gravitational influence of 
the central black hole dominates,
where 
$c_{s}$, and 
$kT_{e}$ are 
the sound speed and
the temperature of
the surrounding hot atmosphere 
\citep[e.g.,][]{russell15}.
For NGC~315, the temperature of 
the hot atmosphere within 1~arcsec has 
$kT_{e}\approx 0.44~{\rm keV}$ which leads
$R_{\rm Bondi}\approx 1.5 \times 10^{6}~r_{g}$,
which is significantly larger than $Z_{\rm brk}$.

Next, we estimate magnetic field strength on the horizon scale ($B_{H}$) by assuming BZ process is in action at the jet base of NGC~315.
The  BZ power and 
the ratio of the angular velocity of the magnetic field line and the event horizon can be given by
\begin{align}\label{eq:LBZ}
 L_{\rm BZ}&\approx
7.5\times 10^{45}\chi_{-2} \frac{\Omega_{F}(\Omega_{H}-\Omega_{F})}{\Omega_{H}^{2}} 
\left(\frac{B_{H}}{10^{3}~{\rm G}}\right)^{2}
\left(\frac{r_{H}}{10^{15}~{\rm cm}}\right)^{2}
~{\rm erg~s^{-1}} , 
\end{align}
and
\begin{align}\label{eq:OmegaF/OmegaH}
\frac{\Omega_{F}}{\Omega_{H}} &= 
\frac{2r_{g}}{r_{\rm lc}} \left(\frac{1+\sqrt{1-a_{*}^{2}}}{a_{*}} \right)
\end{align}
where
$\chi = \int^{\theta_{H}}_{0}   \sin^{3}\theta d\theta
=\frac{2}{3} -\frac{3}{4}\cos\theta_{H} +\frac{1}{12}\cos3\theta_{H}$ \citep[e.g.,][]{Beskin00, takahashi21}
and $r_{H}=r_{g}(1+\sqrt{1-a_{*}^{2}})$ is the outer horizon radius of the black hole and 
$a_{*}\equiv a/M_{\bullet}$ is the dimensionless spin parameter.
For simplicity, $r_{H}$ for $a_{*}\approx 1$ is shown here.
Here we normalized $\chi$ with a typical value 
suggested at the jet base 
$\chi_{-2}=\chi/10^{-2}$ \citep[e.g.,][]{Beskin00, Tchekhovskoy11}. 
From Eqs.~(\ref{eq:LBZ}) and (\ref{eq:OmegaF/OmegaH}), 
it can be readily seen that
$a_{*}$ and $r_{\rm lc}$ are the two parameters
that govern the value of $L_{\rm BZ}$.

Note that there is uncertainty in the value of 
$\chi$ due to $\theta_{H}$.
At a glance, it seems natural to substitute $\theta_{H}\approx \theta_{0}$ with the 4-6 degrees measured at the jet downstream. However, this value leads to $\chi\sim O(10^{-4})$, causing a problem of unphysically large BH being required through Eq. (10). 
This may be probably due to the rapid change in the magnetic field shape near the horizon scale, deviating significantly from the smooth parabolic shape (assumed in TT03).
Indeed, GRMHD simulations for MAD state have shown a tendency for the magnetic field to abruptly change to a larger opening angle near the BH, penetrating the horizon \citep[e.g.,][]{Tchekhovskoy11}. 
Therefore, integrating only within the narrow range of $\theta_{0}=4-6$ degrees on the event horizon significantly underestimate the magnetic flux threading the event horizon. To compromise this problem, we set $\theta_{H}$ as approximately 1 radian in light of GRMHD simulations. 
In the case of M87, actual 86~GHz VLBI image shows
of order of 1 radian half opening angle \citep{hada18,Kim18}. 
Hence, the choice of $\theta_{H}$ did not matter \citep{kino22}.
On the contrary, the jet base of NGC~315 has not yet been spatially resolved to that depth. Hence, we assume $\theta_{H}\approx 1$ radian.

\begin{figure} 
\includegraphics
[width=20cm]
{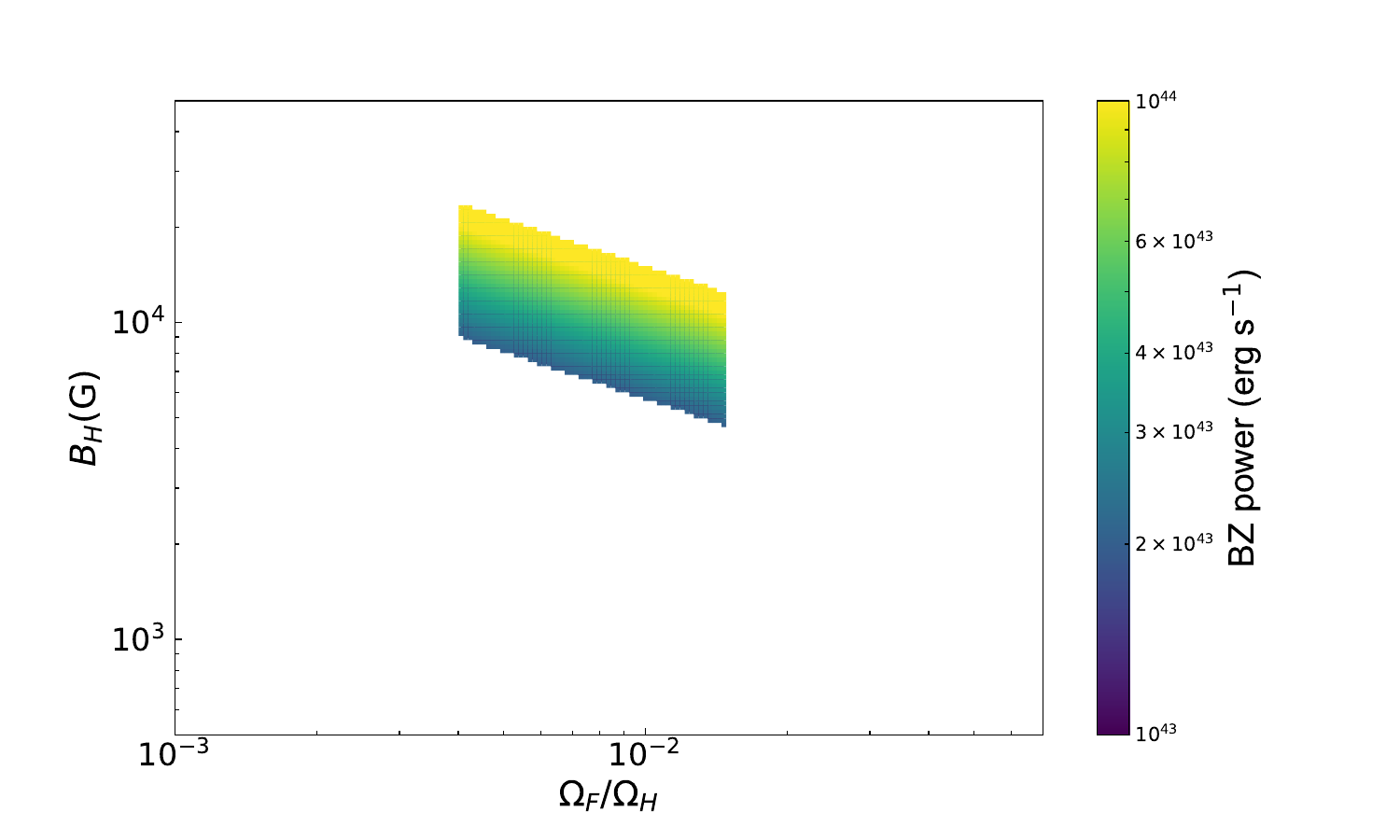}
\caption
{Estimate of  the magnetic field strength 
threading the event horizon ($B_{H}$). 
The horizontal axis shows the ratio $\Omega_{F}/\Omega_{H}$. 
The estimate of $B_{H}$ is done by equating 
$L_{\rm BZ}$ to the total jet power of NGC~315, which is
suggested as 
$2\times 10^{43}~{\rm erg~s^{-1}}\lesssim  L_{j} 
\lesssim 1 \times 10^{44}~{\rm erg~s^{-1}}$ in literature. 
The color bar shows the corresponding $L_{\rm BZ}$.}
\label{fig:BZpower}
\end{figure}

Following the recent GRMHD simulations of highly magnetized jets
\citep[e.g.,][]{Porth19, Ripperda19},
we set the value of the magnetic field threading angle as 
$\theta_{H}= 1$ radian and it leads to $\chi = 0.18$.
Here, we assume the allowed range of the black hole spin as 
$0.5\lesssim a_{*} \lesssim 1$
\citep[e.g.,][]{Zamaninasab14,Nakamura18}
since too small spin is not able to explain the required jet power.
By combining the estimated $\Omega_{F}$ and assumed $\Omega_{H}$
together with the assumption of BZ-driven jet i.e.,  $L_{\rm BZ}\approx L_{j}$,
we obtain the allowed range of   $\Omega_{F}/\Omega_{H}$ and the corresponding 
$B_{H}$ as follows:
\begin{align}
    4.0\times 10^{-3}
    \lesssim \frac{\Omega_{F}}{\Omega_{H}}  \lesssim 1.5\times 10^{-2} , 
    \quad
     5\times 10^{3}~{\rm G}
     \lesssim B_{H}  
     \lesssim 2\times 10^{4}~{\rm G},
\end{align}
Figure~\ref{fig:BZpower} presents the estimated range of $B_{H}$. The lower end of $B_{H}$ is comparable to that estimated in the previous work \citet[][]{ricci22}.
We stress that one of the uniqueness of our analysis is the estimation of $\Omega_{F}$ based on the observed jet velocity profile.

\subsection{Constraining $B(z)$ at the jet downstream} 
\label{sec:Bjet}

Following the method proposed by \citet{Ro23}, 
here we will constrain the physical quantities of the NGC~315 jet from the observed spectral index distribution.
We use the data obtained on January 05th, 2020 \citep[project code: BP243;][]{park21} to investigate the spectral index distribution of the NGC 315. In particular, we use the higher frequency pair (15\,GHz and 22\,GHz) in these multi-frequency observations, since this is less effective by the synchrotron self-absorption.
The information of data is summarized in Table 1 of \citet{park21}. 
To avoid artificial spectral steepening, we used the visibility data with a common $(u, v)$-range of 8 to 445 $M\lambda$, where $\lambda$ is the wavelength of the corresponding radio wave.
We convolved the two images into a circular beam with a radius of 0.9 mas. We also correct the core shift between two frequencies. According to \citet{park21}, the position difference between the 15 GHz core and the 22 GHz core is $0.08\pm0.06$ mas in right ascension and $0.07\pm0.05$ mas in declination.

\begin{figure} 
\includegraphics
[width=18cm]
{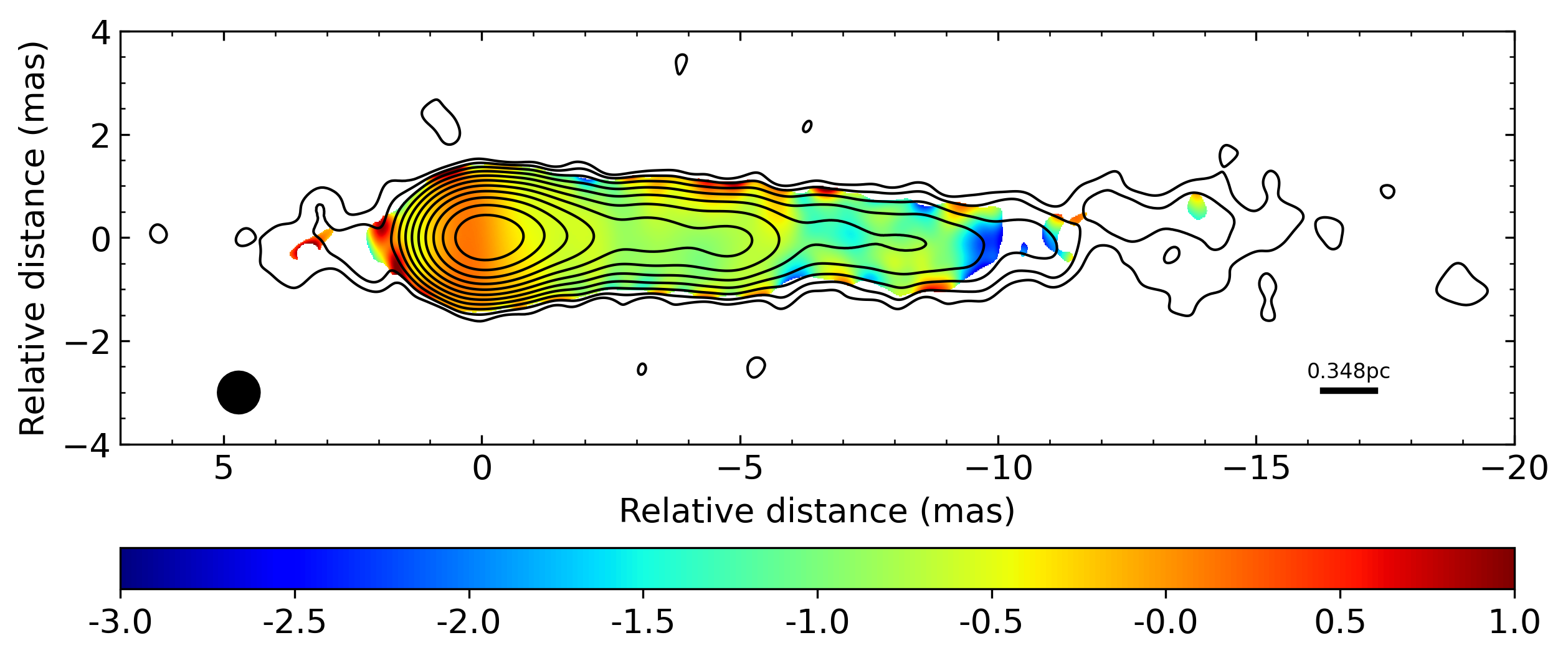}
\caption
{
The spectral index map between 15~GHz and 22~GHz.
The observations at each frequency were quasi-simultaneously conducted
on 2020 January 5th \citep{park21}.
The images at both frequencies 
are restored with the circular Gaussian beam of $0.9$~mas shown at the left-bottom corner.
}
\label{fig:spix_map2}
\end{figure}

Figure \ref{fig:spix_map2} shows a spectral index map of NGC 315 between 15\,GHz and 22\,GHz. We obtain spectral index values up to $\sim$10 mas from the radio core. It also shows that the core has a relatively flat spectrum while the jet has a steep spectrum, which has already been reported in previous studies \citep{park21, ricci22}.
Using this spectral index map, we constructed the spectral index distribution of the jet by taking the weighted average of the spectral indices in the direction perpendicular to the jet as it moves down the jet pixel by pixel. 
When calculating the weighted mean, we used pixel values located within one beam size from the jet axis ($\pm$0.9 mas). 
The error is estimated by multiplying the formal error of the weighted mean by $\sqrt{n}$, where $n$ is the number of the pixel, to compensate for the biases due to correlations between pixel values.

\begin{figure} 
\includegraphics
[width=9cm]
{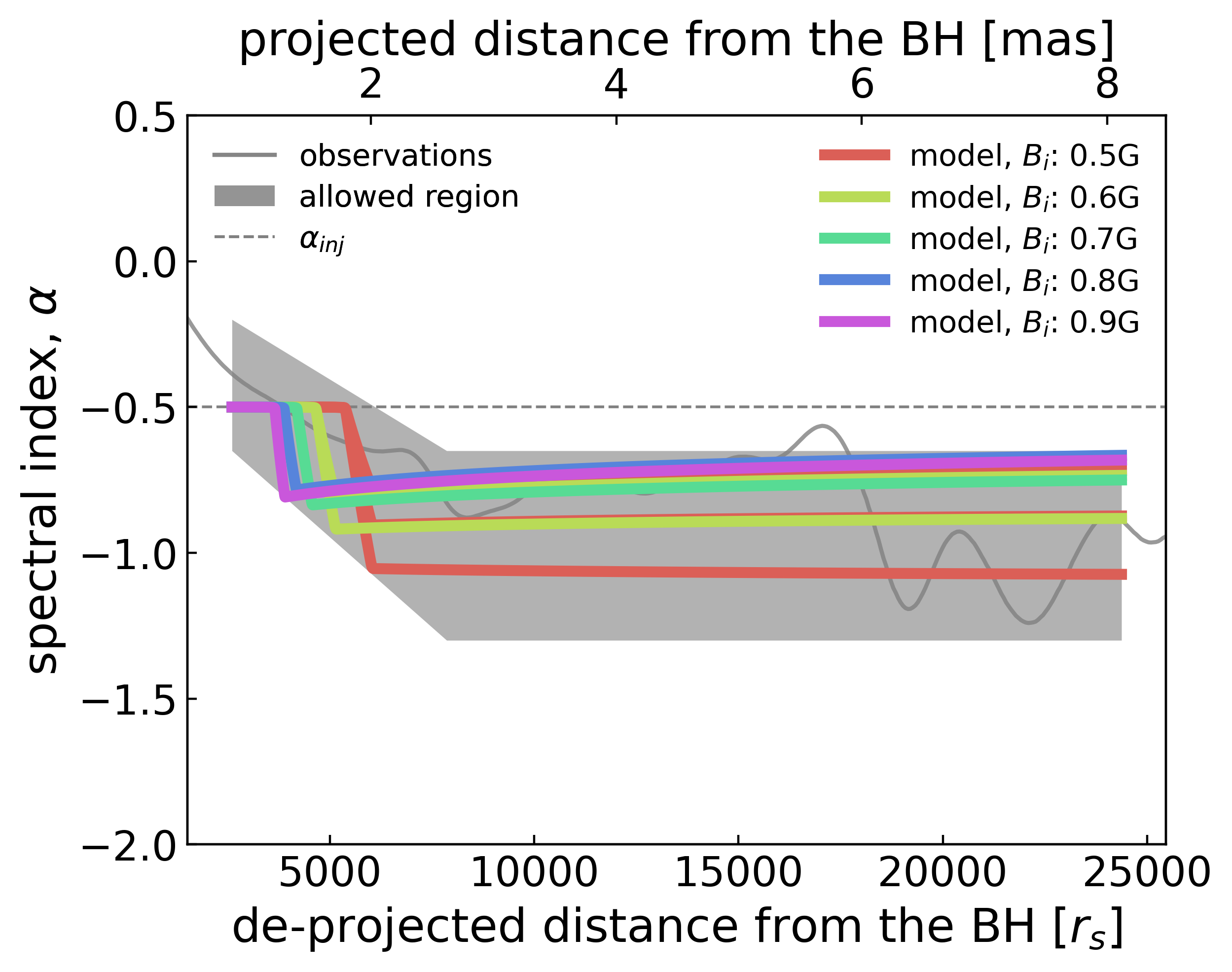}
\includegraphics
[width=9cm]
{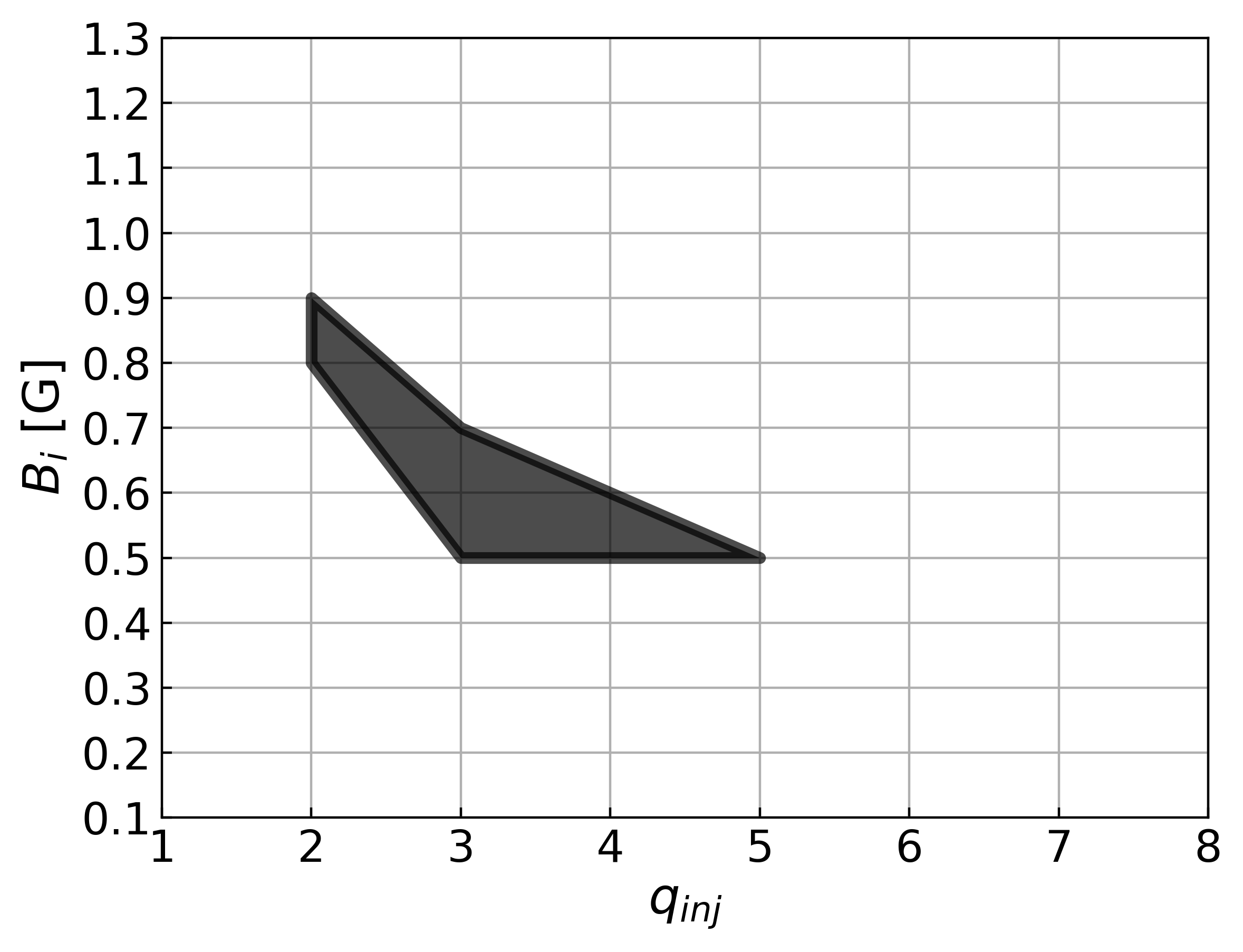}
\caption
{
The radial profile of spectral index between 15~GHz and 22~GHz.
In the range from 0.75 to 2.5~mas, 
the spectral index value decreases from 0 to -1.
In the range from 2.5 to 8~mas, 
the spectral index remains about $-1$ with fluctuations.
The systematic error in the spectral index value 
is conservatively estimated as $10\%$. 
}
\label{fig:spix_distribution2}
\end{figure}

Figure~\ref{fig:spix_distribution2} summarizes the results of the 
spectral index analysis for the NGC~315 jet.
In the left panel, the gray solid line shows the radial distributions of the spectral index between 15~GHz and 22~GHz ($\alpha_{\mathrm{15-22\,GHz}}$) as a function of de-projected distance from the SMBH in units of $r_{g}$.
The spectral distribution out to 8 mas (corresponding to 49,000~$r_{g}$ in the de-projected distance) from the central black hole is obtained.
A steepening of the spectral index is seen until $\sim$2.5 mas, after which it is constant. This behaviour is qualitatively 
same to the one observed in the M87 jet \citep{Ro23}.
By applying the spectral index distribution model 
described in \S~\ref{sec:syn-model},
here we constrain $B_{i}$ and $q_{\rm inj}$.
Using $B_{i}$,
the $z$-profile of the magnetic field strengths $B(z)$  
over the region 0.75 -- 8~mas
(corresponding to 5,200 -- 49,000 $r_{g}$ in the de-projected distance)
is given by
\begin{align}    
B(z)=  B_{i}
\left(\frac{z}{5.2\times 10^{3}~r_{g}}\right)^{-0.88} ,
\end{align}
where the relations of
$B(z)\propto 1/R\Gamma$,
$\Gamma \propto z^{0.3}$, and 
$R \propto z^{0.58}$ are used \citep{Zamaninasab14, park21}.
To compare the model and the observations, 
the allowed region is defined based on the observational spectral index distributions 
with three parameters: 
(1) The spectral index $\alpha_{i}$ at $z_{i}$, 
(2) the location of the end of the steepening region ($z_{f,s}$), and 
(3) the spectral index range after the steepening region ($\alpha_{f}$).
These parameters for the NGC~315 jet are
$-0.65 <\alpha_{f} < -0.2$.
$z_{s,f} = 2.5$~mas,
$-1.3 <\alpha_{f} <-0.65$.
We then create many modeled spectral index distributions for different values of the parameters $B_{i}$ and $q_{\rm inj}$. 
The field strength $B_{i}$ 
was set from 0.1 G to 1.3 G 
in 0.1 G increment, and the injection index is set as $ 1 \le q_{\rm inj} \le 8$.
In all cases, we assumed a constant slope of the nonthermal electron injection function as $p_{\rm inj} = -2.0$ (i.e., $\alpha_{\rm inj} = -0.5$, 
horizontal dashed line in Figure \ref{fig:spix_distribution2}, left).
We limited the parameters by excluding models that exist outside the allowed range. 
Thus, applying the synchrotron-emitting jet model of \citet{Ro23}
to the VLBI observation data, we finally obtain the allowed range as $0.5~{\rm G}\leq B_{i}\leq 0.9~{\rm G}$ together with $2\leq q_{\rm inj}\leq 5$ (Figure \ref{fig:spix_distribution2}, right).

We also applied our model to the spectral index distribution between 22\,GHz and 43\,GHz
in the literature 
\citep[$\alpha_{\mathrm{22-43\,GHz}}$; Figure 2 of][]{ricci22}. 
From this, we examine the spectrum of the region between 0.22 and 2.2 mas ($\sim$1,600 -- $\sim$13,500 $r_{g}$ in de-projected distance) from the black hole. 
However, $\alpha_{\mathrm{22-43\,GHz}}$ shows a fast steepening at $\lesssim$ 0.4 mas, and the amount of steepening is larger than $\sim 0.5$, suggesting so-called fast-cooling regime for nonthermal electrons \citep{sari98}.
Therefore, the spectral index between 22\,GHz and 43\,GHz gives the lower limit of $B(z)$ with $B_{i}=0.8$~G at $z=1.6\times 10^{3}~r_{g}$.

\subsection{Magnetic field strengths along the jet}

\begin{figure} 
\includegraphics
[width=18cm]
{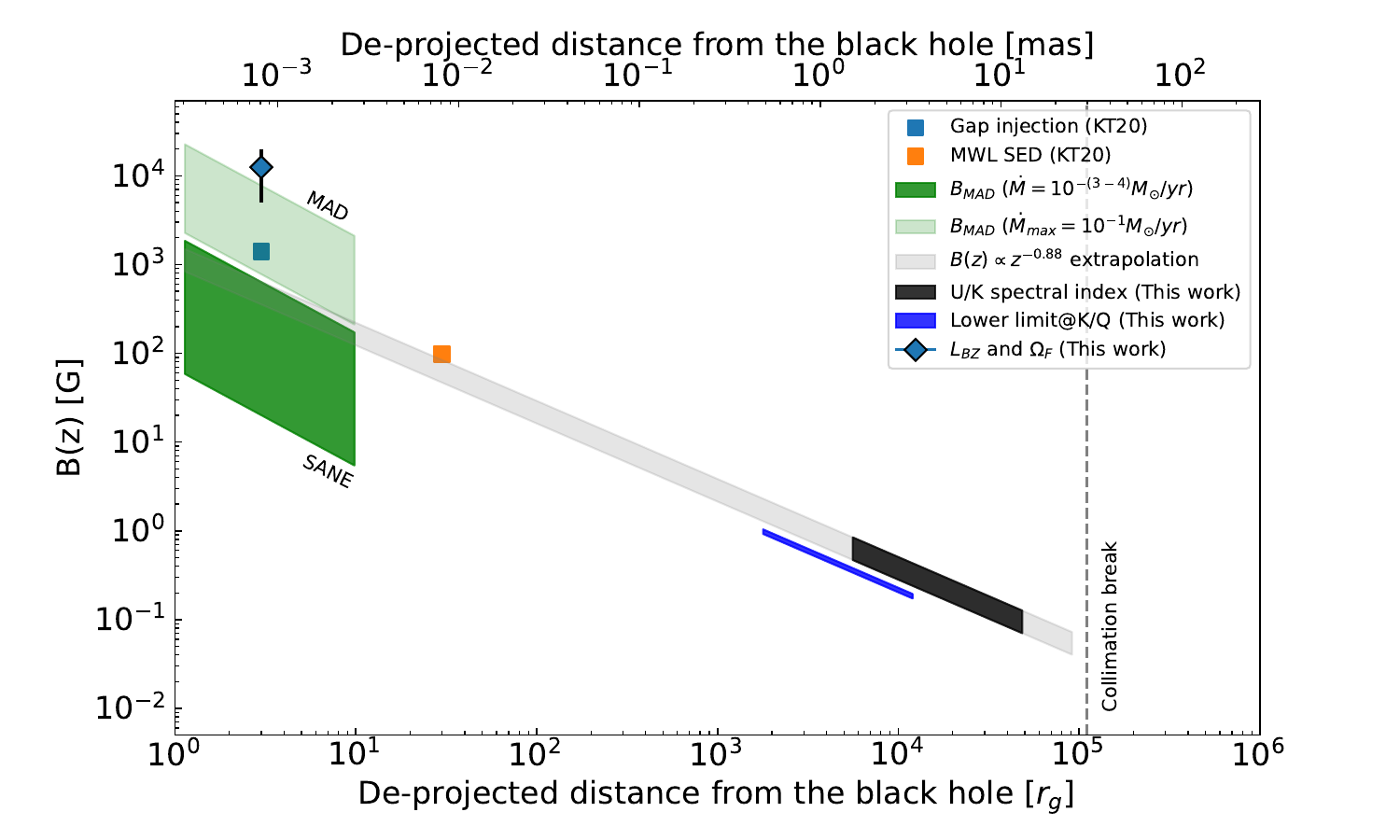}
\caption
{
Mapping the distribution
of the magnetic field strength along the NGC~315 jet.
The black region shows $B(z)$ estimated by our spectral index map analysis made at 15 GHz and 22 GHz for the jet downstream region of
$5.2 \times 10^{3}~r_{g}\lesssim z \lesssim 4.9 \times 10^{4}~r_{g}$. 
The gray-shaded region is the extrapolation of the black area with the $B(z)\propto z^{-0.88}$.
The green-shaded region shows the estimated $B_{\rm MAD}$ in the range from $r_{g} \le z \le 10~r_{g}$ for $10^{-4}~M_{\odot}~{\rm yr^{-1}}\lesssim \dot{M} \lesssim  10^{-3}~M_{\odot}~{\rm yr^{-1}}$.
In addition, $B_{\rm MAD}$ for $ \dot{M}_{\rm max}$ shown (thin-green-shaded region) where $ B_{\rm MAD}$ is
defined in \citet{ricci22}.
The MAD regime realizes for $B(z)\gtrsim B_{\rm MAD}$,
while the SANE regime is expected for $B(z)\lesssim B_{\rm MAD}$.
At the horizon scale, $B_{H}$ estimated in this work is plotted, which clearly shows $B_{H} > B_{\rm MAD}$.
Independent estimates by \citet{KT20} are also plotted
labelled as KT20.}
\label{fig:B_profile}
\end{figure}

In Figure~\ref{fig:B_profile}, 
we show the estimated $B_{H}$ and $B(z)$ along the NGC~315 jet.
In the subsection \ref{sec:BH}, 
we have derived $B_{H}$ based on the constraints from $L_{\rm BZ}$ and the estimated $\Omega_{F}$ values and it resides 
in the range 
$ 5\times 10^{3}~{\rm G} \lesssim B_{H}  
\lesssim 2\times 10^{4}~{\rm G}$ and this is plotted
at $z=r_{g}$.
In the subsection \ref{sec:Bjet}, 
we have also constrained the field strength
at the jet downstream region based on the radial distribution of the spectral index between 15 and 22 GHz, and the black area is the corresponding allowed range of $B(z)$ 
in the range of $0.75~{\rm mas} \lesssim z \lesssim 8~{\rm mas}$. 
The gray-shaded region is the extrapolation of the filled-black region.
The lower limit of $B(z)$ (blue-colored) is also appended, which is estimated from the radial distribution of the spectral index between 22 and 43 GHz.
In addition, other independent estimations of the magnetic field strengths are also plotted. \citet{KT20} explored multi-wavelength emission spectrum from NGC~315  and suggested that the observed $\gamma$-ray from NGC~315 is well explained by
the hadronic emission from MAD.
\footnote{
\citet{Igumenshchev03} found that, given the right initial
conditions in MHD simulations, magnetic fields can become dynamically important in BH accretion flows, 
to the extent that they impede the inward motion of gas and create a ‘magnetically arrested disc’ and this is named as MAD \citep{Narayan03}, see also \citet{Bisnovatyi-Kogan74}. 
Hot accretion flows in the MAD regime can
launch powerful jets.}

To understand physical properties 
of the accretion flow and the jet base,
we make an order estimation of  the magnetic field strengths in a MAD regime as a reference value
by using the literature value of $\dot{M}$ for NGC~315. 
In light of \citet{Narayan03},
the magnetic field supports the plasma against 
the gravitational pull by the black hole. 
The strength of the magnetic field 
$B_{\rm MAD}$ that satisfies the force balance between the gravitational pull and the magnetic support can be given by
$B_{\rm MAD}\approx ( 2M_{\bullet}\dot{M}/3 r^{3}v_{r,{\rm disk}} h_{\rm disk}/r_{\rm disk})^{1/2}$
where 
$v_{r,{\rm disk}}$,
$ h_{\rm disk}$, and
$r_{\rm disk}$ are 
the radial velocity of the accreting matter,
the height of the accretion disk, and 
the radial distance in the accretion disk,
respectively
\citep[][and references therein]{ricci22}.
The estimated $ B_{\rm MAD}$ value (the green-shaded region in Figure~\ref{fig:B_profile}) indicates the baseline that the SANE region will realize if the estimated value is below that value.
Here we assume a scale height of the accretion disk  $h_{\rm disk}/r_{\rm disk} \approx 1$ and a medium value of $v_{r,{\rm disk}}=0.1c$ which is 
between the value assumed in \citet{Narayan03}
(i.e., $v_{r,{\rm disk}} \approx 10^{-2}c$) and the value assumed in \citet{ricci22} (i.e, $v_{r,{\rm disk}}\approx c$).
Unfortunately, the uncertainty of $\dot{M}$ is so large that the divided region is fairly thick.
Therefore, the values of $h_{\rm disk}/r_{\rm disk}$ and 
$v_{r,{\rm disk}}$ does not affect $B_{\rm MAD}$ significantly. 
As shown in Fig.~\ref{fig:B_profile}, we find that 
the extrapolated region (gray-colored) is consistent with MAD regime.  The estimated $B_{H}$ in sub-section \ref{sec:BH} also suggests that the accretion flow of NGC~315 is in a MAD regime.
This claim agrees with the previous study of \citet{ricci22}.
In addition, the magnetic field strengths estimated 
in \citet{KT20} also reside in the MAD regime.

\section{Discussion}

\subsection{A note on a small $\Omega_{F}/\Omega_{H}$}\label{sec:smallOmegaF}

Previous works develop an analogy of direct current (DC) circuit analysis for power flow through the black hole magnetosphere, known as the membrane paradigm of black holes \citep[][and references therein]{thorne86}.
In this context, we consider a meaning of a small $\Omega_{F}/\Omega_{H}\sim 10^{-2}$
indicated in this work.
A plasma-filled (i.e., non-vacuum) black-hole magnetosphere resembles a DC circuit. The magnetized rotating black hole acts as a battery driving the current ($I$). Strong magnetic field lines threading the event horizon function like current-carrying wires, allowing charged particles to slide along them. These magnetic wires connect the event horizon to a distant, weakly-magnetized load region, where the current returns to the black hole battery.
This DC circuit analysis 
between a certain two magnetic-field surfaces 
can give the relation among,
$\Omega_{H}$,
$\Omega_{F}$,
the resistance across the event horizon ($\Delta R_{\rm H}$),and 
the load region ($\Delta R_{\rm load}$) 
as follows:
\begin{align}
\frac{\Omega_{H}-\Omega_{F}}{\Omega_{F}}=
\frac{ \Delta R_{\rm Load}}{\Delta R_{\rm H}}=\frac{ \Delta V_{\rm Load}}{ \Delta V_{\rm H}}  ,
\end{align}
where 
$\Delta V_{\rm H}=I\Delta R_{\rm H}$, and
$\Delta V_{\rm Load}=I\Delta R_{\rm Load}$, are
the voltage drop at the event horizon and
that of the load region, respectively
\citep[see Section IV in the textbook of][]{thorne86}.
If assuming the power into the load region is maximized,
the analogy of a laboratory DC circuit suggests 
the impedance matching of 
$\Delta Z_{\rm H} \approx \Delta Z_{\rm Load}$ 
\citep{macdonald82},
leading to $\Omega_{F} \approx \Omega_{H}/2$.
GRMHD simulations have also suggested that the ratio
$\Omega_{F}/\Omega_{H}$ fits in the similar range of $0.3-0.5$ \citep[e.g.,][]{McKinney07,Penna13}. 
Therefore, $\Omega_{F}/\Omega_{H} \sim 10^{-2}$ indicated 
in this work may correspond to an impedance mismatching.
In a DC circuit, an impedance mismatching causes energy transmission losses.
Therefore, to compensate for this lost energy,
a relatively large $B_{H}$ in $L_{\rm BZ}$ is needed in this work
to meet the required $L_{j}$ for NGC~315.
Although detailed physical processes  in the load regions in astropysical AGN jet system are unclear, it seems likely that the electric fields of the load region will accelerate charged particles to high energies \citep{thorne86}.



\subsection{Magnetic flux threading the event horizon of
in NGC~315}

Magnetic flux threading the event horizon ($\Phi_{\rm BH}$) is one of the key parameters that characterizes the jet launching \citep[e.g.,][]{Zamaninasab14}.
Although the dimensionless magnetic flux threading one hemisphere of the event horizon ($\phi_{\rm BH} = 
\Phi_{\rm BH}/\sqrt{\dot{M}r_{g}^{2}c}$
where $\Phi_{\rm BH}\approx B_{H} r_{g}^{2}$ is 
the magnetic flux threading the black hole)
is actively discussed in GRMHD simulations
\citep[e.g.,][]{Tchekhovskoy11,narayan22},
the strength of $\phi_{\rm BH}$ is highly uncertain
in actual jet objects.
Therefore, it is worthwhile to estimate a typical 
$\phi_{\rm BH}$ in order to compare GRMHD simulations
and observations.
Using the estimated $B_{H}$, we can 
make the order estimation of 
$\phi_{\rm BH}$ (in Gaussian-cgs units) for NGC315
as follows:
\begin{align}
    \phi_{\rm BH}=
\frac{\Phi_{\rm BH}}{\sqrt{\dot{M}r_{g}^{2}c}} 
\approx 77
\left(\frac{B_{H}}{1\times 10^{4}~{\rm G}}\right)
\left(\frac{M_{\bullet}}{1.6\times 10^{9}~M_{\odot}}\right)
\left(\frac{\dot{M}}{5\times 10^{-4}~ M_{\odot}~{\rm yr^{-1}}}\right)^{-1/2}  ,
\end{align}
or equivalently
\begin{align}\label{eq:Phi_BH}
    \Phi_{\rm BH} \approx B_{H} r_{g}^{2} 
        &\approx 5.6 \times 10^{32} 
        \left(\frac{\phi_{\rm BH}}{77} \right) 
        \left(\frac{M_{\bullet}}{1.6\times10^{9} M_{\odot}}\right)
        \left(\frac{\dot{M}}{5\times 10^{-4}~M_{\odot}~{\rm  yr}} \right)^{1/2}  \, {\rm G\, cm^{2}} .
\end{align}
%
Since it is generally difficult to estimate $B_{H}$,
\citet{Zamaninasab14} suggested an alternative way  where assuming poloidal magnetic flux threading parsec-scale jets $\Phi_{\rm jet}$ is comparable to  $\Phi_{\rm BH}$.
Using the relation of $\Phi_{\rm BH} \sim \Phi_{\rm jet}$,
they indicated that the jet-launching regions of AGN jets are threaded by dynamically important fields, which obstruct gas infall, compress the accretion disk vertically, slow down the disk rotation by carrying away its angular momentum in an outflow and determine the direction of jets.
Our estimated value for NGC~315 shown in Eq.~(\ref{eq:Phi_BH}) 
agrees with the result of \citep{Zamaninasab14}.

In addition, it is also worth checking the value of 
the efficiency with which the black hole generates the jet power ($\eta_{\rm BZ}$) as the ratio of $L_{\rm BZ}$ 
to the inflow rate at which rest mass energy flows into the black hole \citep{Tchekhovskoy11}.
Together with the assumption of $L_{\rm BZ}\approx L_{j}$,
the value $\eta_{\rm BZ}$ for NGC~315 is estimated as
\begin{align}\label{eq:eta_BZ}
    \eta_{\rm BZ} = \frac{L_{\rm BZ}}{\dot{M}c^{2}} \approx
    3.5
\left(\frac{L_{j}}{1\times 10^{44}~{\rm erg~s^{-1}}}\right)
\left(\frac{\dot{M}}{5\times 10^{-4}~ M_{\odot}~{\rm yr^{-1}}}\right)^{-1}   .
\end{align}
The estimation of $ \eta_{\rm BZ} \gtrsim 100~\%$ 
implies that the NGC~315 jet carries away more energy than
the entire rest mass energy of the accreted gas,
suggesting the need of 
extracting the rotational energy of the black hole.
The estimated order of $\eta_{\rm BZ}$ in Eq.~(\ref{eq:eta_BZ}) can be comparable to the maximal value of $\eta_{\rm BZ} \sim 300~\%$
seen in GRMHD simulations \citep[e.g.,][]{Tchekhovskoy11, narayan22}.

\subsection{Comparison with NGC~315 and M87}

\subsubsection{Relation of $\phi_{\rm BH}$
and $L_{j}/L_{\rm Edd}$}\label{sec:phiBH-Lj}

In light of recent GRMHD numerical simulations, as the value of $\phi_{\rm BH}$ increases, the accretion flow mode 
switches from SANE to MAD. 
Based on this knowledge, an ansatz emerges that in actual jet objects, the larger the $\phi_{\rm BH}$ value, the more powerful the jets may be produced. 
However, the estimation and discussion of $\phi_{\rm BH}$  (or $\Phi_{\rm BH}$) value in actual jet objects have hardly been conducted. 
In order to address this  ansatz, 
it would be worth comparing 
the range of estimated $\phi_{\rm BH}$ values 
and $L_{j}/L_{\rm Edd}$ for NGC 315 and M87
even although there are only two sources treated here.
\citet{Zamaninasab14} showed a positive correlation between $\phi_{\rm BH}$ and the square of the luminosity of the accretion disk. 
This also appears to support the positive correlation between $\phi_{\rm BH}$ and jet power, since it is widely known that $L_{j}$ has a positive correlation with the luminosity of the accretion disk \citep[e.g.,][]{rawlings91, ghisellini14}.

\begin{figure} 
\includegraphics
[width=18cm]
{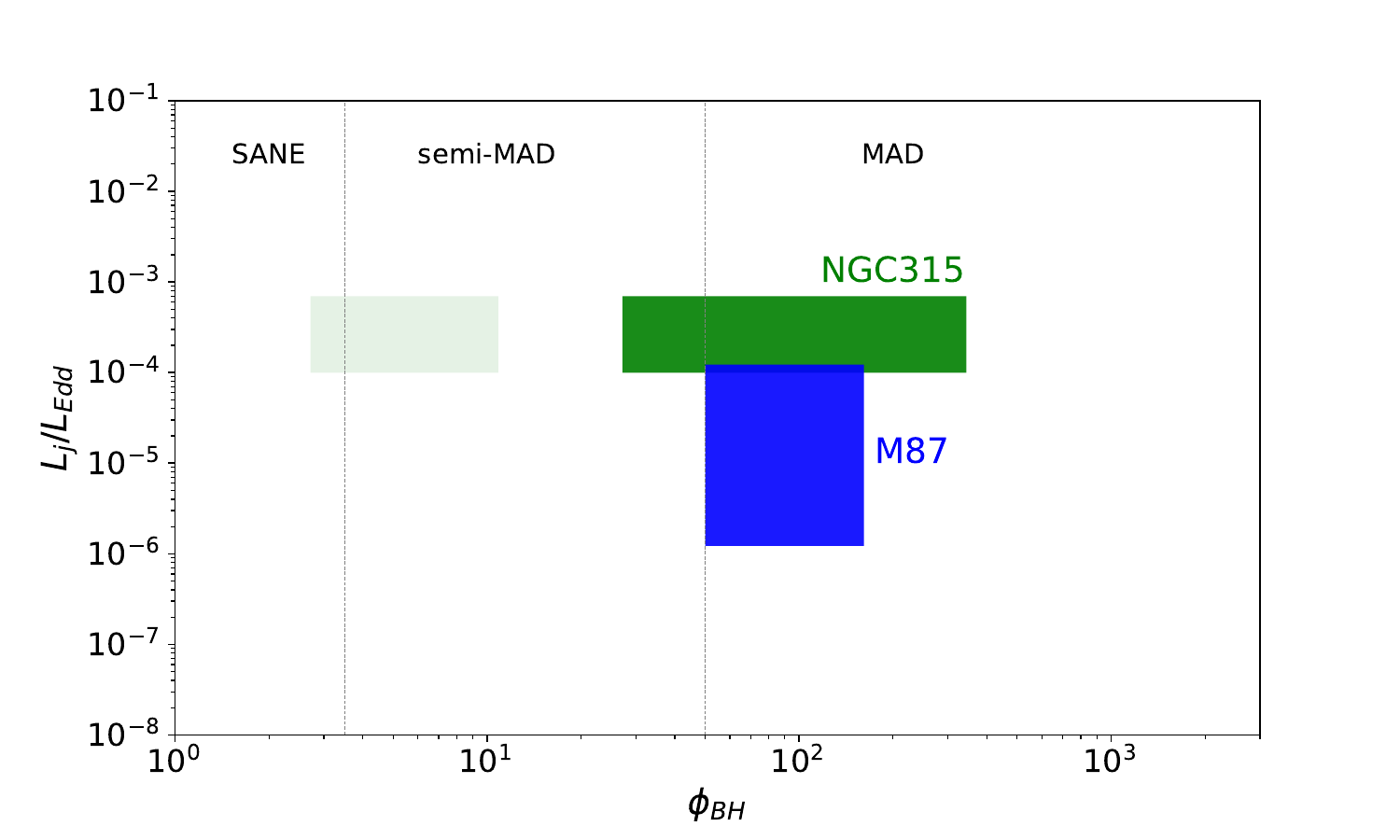}
\caption
{
Estimated allowed ranges of NGC~315 (green) and M87 (blue) in $\phi_{\rm BH}$ (Gaussian-cgs units)
and $L_{j}/L_{\rm Edd}$ plane.
According to \citet{EHTC8}, the case where $\phi_{\rm BH} \gtrsim 50$ is referred to as MAD, while $\phi_{\rm BH} \sim \sqrt{4\pi}$ is referred to as SANE. In this paper, we define the range $3.5 \lesssim \phi_{\rm BH} \lesssim 50$ as semi-MAD.
For M87, the lower limit of $\phi_{\rm BH}\approx 50$ is adopted based on \citet{EHTC8}, while the upper limit 
of $\phi_{\rm BH}$ is estimated based on \citet{kino22}.}
\label{fig:phi_Lj}
\end{figure}

In Figure~\ref{fig:phi_Lj}, we  investigate the relationship between $\phi_{\rm BH}$ and $L_{j}/L_{\rm Edd}$ for NGC~315 and M87. 
According to \citet{EHTC8}, the range $\phi_{\rm BH} \gtrsim 50$ is classified as MAD, while the range below $\phi_{\rm BH} \sim \sqrt{4\pi}$ is classified as SANE. 
The range in between 
$\sqrt{4\pi} \lesssim \phi_{\rm BH} \lesssim 50$ is defined as semi-MAD in this work.
Although the estimation of the $B_{H}$ value depends on the adopted $L_{j}$, 
the estimated $\phi_{\rm BH}$ suggests
that NGC~315 generally aligns with a MAD regime
unless an extremely high mass accretion
rate, $\dot{M}_{\rm max}$, is realized.
However, discussing a correlation between $\phi_{\rm BH}$ and $L_{j}/L_{\rm Edd}$ using 
Figure~\ref{fig:phi_Lj} is challenging due to significant uncertainties in $\phi_{\rm BH}$ caused by $\dot{M}$ and $L_{j}$. In particular, $\dot{M}$ introduces a large uncertainty in $\phi_{\rm BH}$.
Nonetheless, it is important to explore 
the relation between $\phi_{\rm BH}$ and $L_{j}/L_{\rm Edd}$ in terms of exploring BZ77.
To further test BZ77, it is essential to narrow the allowed range of $\dot{M}$.
In the case of M87, polarimetric observation 
by the EHT significantly narrowed the allowed region of  $\dot{M}$ \citep{EHTC8}.
Following this, we expect that future higher sensitivity and higher resolution VLBI observations, such as the next generation EHT (ngEHT) and 
the Black Hole Explorer (BHEX) \citep[e.g.,][]{Kurczynski22,johnson23,doeleman23},
will similarly narrow down the $\dot{M}$ in NGC315.
From a theoretical standpoint, existing GRMHD numerical experiments have not sufficiently explored the parameter range where $\phi_{\rm BH}>100$. 
Investigating a correlation between $\phi_{\rm BH}$ and $L_{\rm  j}/L_{\rm Edd}$ accross a wide range will be an important next step in understanding the jet launching mechanism.


\subsubsection{Implication: What determines the location of 
the $Z_{\rm brk}$?}



Although there are mounting observational measurements of $Z_{\rm brk}$ for various AGN jets
\citep[e.g.,][] {Asada12,Hada13, tseng16,nakahara18,algaba19,nakahara20,kovalev20,boccardi21, yi24},
what determines the location of the $Z_{\rm brk}$ is still an open issue. There are two possible scenarios.
One scenario suggests that the location of the $Z_{\rm brk}$ is associated with differences in the central engine and/or jet's intrinsic magnetic fields \citep{potter15,nokhrina19,kovalev20,nokhrina20},
while the other scenario proposes that
the surrounding matter around the jet determines of the location of the $Z_{\rm brk}$
\citep[e.g.,][]{park21,Okino22,rohoza24}.
NGC~315 seems to be decelerated down to a non-relativistic speed at $Z_{\rm brk}$ which is much closer than $R_{\rm Bondi}$ \citep{park21}. 
For M87, the jet apparent speeds become smaller beyond $Z_{\rm brk}\sim R_{\rm Bondi}$ \citep{Park19}
and it can be understood as the M87 jet is decelerated at
$Z_{\rm brk}\sim R_{\rm Bondi}$.
Both NGC~315 and M87 could be understood in terms of the locations of $Z_{\rm brk}$ being determined by the different locations of the surrounding material. However, it is difficult to determine which of the two possible scenarios is realized because the strength of the central engine is not clearly constrained and compared. To address this, we attempt to investigate the relationship between $\phi_{\rm BH}$ and $L_{j}/L_{\rm Edd}$.
However, it cannot be conclusive with the current best data due to the large uncertainty in $\phi_{\rm BH}$  (see Sub-section \ref{sec:phiBH-Lj}). 
But if a positive correlation between $\phi_{\rm BH}$ and  $L_{j}/L_{\rm Edd}$ could be found in the future, then
that would support the idea that 
surrounding matter around the jet play a more dominant role for determination of the location of the $Z_{\rm brk}$.
Further investigation is eagerly anticipated.

\section{Summary}

In the present work, we investigate magnetic field strengths along the jet in NGC~315, building upon the earlier work by \citet[][]{kino22}. Before summarizing the results, we briefly address two frequently asked questions and provide their answers.

\begin{itemize}

\item 
Some may consider entrainment as a physical process to explain the slower velocity within $\sim 10^{3}~r_{g}$.
It could be possible to decelerate the flow with intense entrainment with surrounding ambient material after the jet's bulk acceleration reaches its maximum velocity $\Gamma_{max}$ at a distance of about $10^{2}~r_{g}$ from the central engine, as suggested in GRMHD simulations
\citep[e.g.,][Fig. 13]{chatterjee19}. The thermal energy generated during the interaction between the jet and surrounding medium may help re-accelerate the jet.
However, the numerical simulation by \citet{ricci23}, which investigated thermal acceleration, did not show sufficient deceleration, resulting in a significant overestimation the jet velocity within $\sim 10^{3}~r_{g}$. Moreover, such significant deceleration and re-acceleration do not produce a smooth velocity profile but rather a beat-like profile, which is not suggested in the velocity profile of NGC~315.
For these reasons, explaining the velocity profile through the entrainment phenomenon appears challenging.

\item 

Assuming maximal power transfer to the load region, akin to a laboratory DC circuit, suggests impedance matching: $\Delta Z_{\rm H} \approx \Delta Z_{\rm Load}$ \citep{macdonald82}, implying $\Omega_{F} \approx \Omega_{H}/2$. 
As discussed in Section~\ref{sec:smallOmegaF},
GRMHD simulations also indicate $\Omega_{F}/\Omega_{H}$ typically falls in the range of $0.3-0.5$ \citep[e.g.,][]{McKinney07,Penna13}.
Thus, $\Omega_{F}/\Omega_{H} \sim 10^{-2}$ suggested in this study may be understood as impedance mismatching. 
In a DC circuit, such mismatching results in energy transmission losses. To compensate, this study requires a relatively large magnetic field $B_{H}$ in $L_{\rm BZ}$ to achieve the necessary jet power $L_{j}$ for NGC~315.

\end{itemize}

Below, we summarize our findings in the present work.

\begin{itemize}
    \item First, we estimated the angular velocity of rotation in the jet magnetosphere by comparing the measured velocity profile of NGC~315 with the MHD jet model of TT03, following the recent work of \citet{kino22} 
    on M87. 
Then, we find that the model can reproduce the logarithmic feature of the velocity profile with  $c/(800~r_{g})\lesssim \Omega_{F} \lesssim c/(500~r_{g})$, suggesting slowly rotating black hole magnetosphere for NGC~315. By substituting this $\Omega_{F}$ into the jet power predicted by the Blandford-Znajek mechanism, we estimate the magnetic field strength near the event horizon as $5\times 10^{3}~{\rm G}\lesssim B_{H}\lesssim 2\times 10^{4}~{\rm G}$.

    \item 

We then estimate the magnetic-field strengths along the jet by comparing the spectral index distribution obtained from VLBI observations with a synchrotron-emitting jet model based on the recent work of \citet{Ro23}. 
Then we constrain the magnetic field strength at the de-projected distance of $5.2 \times 10^{3}~r_{g}\lesssim z \lesssim 4.9 \times 10^{4}~r_{g}$ to be in the range $0.06~{\rm G}\lesssim B(z)\lesssim 0.9~{\rm G}$.
By combining the obtained field strengths at the event horizon and the downstream section of the jet, we find that the accretion flow at the jet base is consistent with a MAD. 

\item 

Using the estimated $B_{H}$ in this work, 
the dimensionless magnetic field flux threading the event horizon $\phi_{\rm BH}$ is  estimated.
It can be well above $\phi_{\rm BH} > 50$, suggesting MAD regime is realized in NGC~315.
In addition, the efficiency of the jet power production 
($\eta_{\rm BZ}$) for NGC~315 can be comparable with the maximal value of $\eta_{\rm BZ} \sim 300~\%$ seen in GRMHD simulations \citep[e.g.,][]{Tchekhovskoy11, narayan22}.

    \item 
The comparison of $\phi_{\rm BH}$ and $L_{j}/L_{\rm Edd}$  between NGC~315 and M87 has been made.
However, it is difficult to clarify a relation between $\phi_{\rm BH}$ and $L_{j}/L_{\rm Edd}$ because of uncertainties in $\dot{M}$. This causes a large uncertainty in $\phi_{\rm BH}$.
To improve our understanding of BZ77, we expect that future higher sensitivity and higher resolution VLBI observations, such as ngEHT and BHEX \citep[e.g.,][]{Kurczynski22, johnson23, doeleman23}, would narrow down the $\dot{M}$, thereby providing a better constraint on $\phi_{\rm BH}$ in NGC~315.

\end{itemize}


\bigskip
\leftline{\bf \large Acknowledgment}
\medskip

\noindent

We thank the referee for his or her constructive comments and careful reading of the manuscript, which helped improve its overall clarity.
This work was partially supported by
the MEXT/JSPS KAKENHI
(JP17K05439,
JP21H01137,
JP21H04488,
JP22H00157,
JP23H00117, and
JP23K03448).
 This research was also supported by MEXT as “Program for Promoting Researches on the Supercomputer Fugaku” (Toward a unified view of the universe: from large scale structures to planets, JPMXP1020200109) and JICFuS. 

\newpage

\bibliography{ms_r1.bbl}{}




\end{document}